# Emergent Transition for Superconducting Fluctuations in Antiferromagnetic Ruthenocuprates

A. C. Mclaughlin [1] and J. P. Attfield [2]

[1] Department of Chemistry, University of Aberdeen, Meston Walk, Aberdeen AB24 3UE, UK.

[2] Centre for Science at Extreme Conditions and School of Chemistry, University of Edinburgh, King's Buildings, Mayfield Road, Edinburgh EH9 3JZ.

The emergence of carrier-pairing from the electronically inhomogeneous phase of lightly hole-doped copper oxides has been investigated through magnetoresistance measurements on 1222-type ruthenocuprates $RuSr_2(R,Ce)_2Cu_2O_{10-\delta}$, principally with $R$ = Gd, Sm, Nd. A well-defined transition at which superconducting fluctuations emerge is discovered at a remarkably low critical doping, $p_c$ = 0.0084, deep within the antiferromagnetic phase. Systematic variations of the low temperature fluctuation density with doping and cell volume demonstrate the intrinsic nature of the electronic inhomogeneity and provide new support for bosonic models of the superconducting mechanism.

A key issue for understanding high temperature superconductivity in copper oxides (cuprates) is to discover how carrier-pairing emerges from the electronically inhomogeneous region of the phase diagram between the antiferromagnetic insulator (typically at hole doping concentrations $0 < p < 0.02$) and bulk superconducting ($0.05 < p < 0.25$) regimes. Recent measurements have evidenced local pairing fluctuations at dopings just below the onset of bulk superconductivity in $La_{2-x}Sr_xCuO_4$ films [1], and in the pseudogap regime above the critical temperature $T_c$ for higher dopings [2, 3], but their extent and interpretation remain controversial [4, 5]. In this Letter we use magnetoresistance measurements on 1222-type ruthenocuprates [6, 7, 8], to show that superconducting fluctuations emerge from a well-defined transition deep within the antiferromagnetic phase, providing new support for bosonic models of the superconducting mechanism [9, 10, 11, 12].

The resistive response of doped copper oxide planes to a magnetic field is greatly enhanced by the ruthenium oxide layers in 1222-type ruthenocuprates [6, 7, 8]. Ruthenium spins order in a canted antiferromagnetic arrangement at temperatures $T_{Ru} \approx$ 120-180 K above $T_c$ or the ordering temperature for copper spins $T_{Cu}$, so the intrinsic physics of doped $CuO_2$ planes is observed against a background of ordered Ru spins. Applied magnetic fields cant the spins towards ferromagnetic alignment, greatly enhancing the mobility of holes in $CuO_2$ planes. Here we analyse magnetoresistances ($MR = [\rho(H) - \rho(0)]/\rho(0)$ from measurements of electronic resistivity $\rho$ in applied field $H$ and at $H = 0$) for a large number of lightly-doped 1222-ruthenocuprate samples. These are primarily series of $RuSr_2R_{1.8-x}Y_{0.2}Ce_xCu_2O_{10-\delta}$ ($R$ = Gd, Sm, Nd) materials which are stable for doping levels $p = 0.01 - 0.06$; $p$ is calculated from Ce content $x$ and measured oxygen deficiency $\delta$ as $p = (1 - x - 2\delta)/2$ [13]. $MR$ measurements on previously reported 1222 samples with fixed $p = 0.03$ doping [7, 8] are also used. Although grain boundaries contribute significantly to the resistivities of such polycrystalline ceramic samples, other transport measurements such as Seebeck effect and here magnetoresistance are less sensitive and provide information about intrinsic properties.

On cooling from ambient temperature, $MR$ becomes increasingly negative with discontinuities due to spin ordering at $T_{Ru}$, and at $T_{Cu}$ for non-bulk-superconducting samples (Fig. 1). Previous neutron diffraction studies of $R$ = Nd samples confirmed the onsets of long range spin order at these transitions [6]. In the $R$ = Gd series shown in Fig. 1a, samples with $p > 0.04$ are bulk superconductors with transitions to the zero resistance state while less-doped materials are insulating. $MR$ becomes positive and divergent for superconducting samples on cooling through $T_c$ but it is notable that all the insulating $R$ = Gd samples down to the lowest doped $p = 0.014$ material also show a broad, positive upturn, although $MR$ returns towards negative values characteristic of the antiferromagnetic phase on further cooling. The upturn feature evidences traces of superconductivity that we attribute to pair fluctuations in the antiferromagnetic regime. The upturns have a constant onset temperature of $T_{sf}$ = 27 K, close to the maximum $T_c$ of 28 K reported in bulk superconductors from this series [14], demonstrating that the samples are electronically inhomogeneous, but results shown later reveal that the density of low temperature fluctuations varies systematically between samples and is an intrinsic quantity. Magnetoresistances of $R$ = Sm samples give similar results [13] with $T_{sf}$ = 24 K.

Electronic phase diagrams derived from the $MR$(T) data for the R = Gd and Sm systems are shown in Fig. 2. The doping onset of superconductivity varies strongly with $R$, increasing from $p \approx$ 0.04 for Gd to $p$ = 0.055 for Sm to $p$ > 0.06 for Nd where no superconductivity was observed at this upper doping limit [6]. Low temperature $MR$-$H$ variations are shown in Fig. 3a. Negative $MR$ decreasing almost linearly with field at low $p$ is characteristic of the antiferromagnetic phase, and the large positive contribution seen at low fields as doping increases is due to superconducting fluctuations. The increasing contribution of superconducting fluctuations with $p$ is illustrated by plotting $MR/H$ as described in Fig. 3b.

We use values of the magnetoresistance at a constant low temperature (4 K) and high field strength (7 T) to quantify the density of superconducting fluctuations across many 1222 ruthenocuprate samples. A striking result is that $MR_{7T,4K}$ values from the RuSr$_2R_{1.8-x}$Y$_{0.2}$Ce$_x$Cu$_2$O$_{10-\delta}$ ($R$ = Gd, Sm, Nd) samples and many others belonging to different 1222 phase diagrams (e.g. RuSr$_2$(Nd,Tb,Ce)$_2$Cu$_2$O$_{10-\delta}$, RuSr$_2$(Sm,Ce)$_2$Cu$_2$O$_{10-\delta}$, (Ru,Ta)Sr$_2$(Nd,Y,Ce)$_2$Cu$_2$O$_{10-\delta}$; the full list is in Supplemental Material [13]) with a constant $p$ = 0.03 doping, show a linear variation with unit cell volume (Fig. 4a). This demonstrates that the low temperature density of superconducting fluctuations observed in antiferromagnetic samples is an intrinsic quantity and leads to simple $MR$-behaviours as described below.

Detailed analysis of fluctuation contributions to the normal-state magnetoresistances of superconductors is complicated [15] but a simple classical approximation may be used to compare the $MR_{7T,4K}$ values between different lightly-doped samples here. We write the effective scalar resistivity as $\rho = \rho_a/(1 + 3c)$ following Maxwell's expression for a dilute suspension of non-interacting conductive particles (here superconducting fluctuations) of volume concentration $c$ embedded in a matrix of (antiferromagnetic) insulator with resistivity $\rho_a$ [16]. Upper critical fields ~40 T are reported for bulk 1222 ruthenocuprate superconductors [17] but the field scale is greatly reduced for local fluctuations (e.g. to ~10 T in $La_{2-x}Sr_xCuO_4$ films at $p \approx 0.05$ 1), and a 7 T field is sufficient to break all superconducting fluctuations in our $p \leq 0.03$ materials giving $\rho(H) = \rho_a(H)$ and hence;

$$MR = MR_a + 3fc_{max}(p)[MR_a + 1] \qquad (1)$$

The magnetoresistance of the antiferromagnetic matrix in the absence of pair fluctuations, $MR_a$, is negative but the second term gives a positive contribution weighted by the density of superconducting fluctuations $c = fc_{max}(p)$. $f$ represents the fraction ($0 < f < 1$) of the maximum concentration $c_{max}(p)$ possible for doping $p$.

The strong linear decrease in $MR_{7T,4K}$ with unit cell volume over the range 421 to 426 $Å^3$ in Fig. 4a is consistent with Eq. 1 assuming the fraction of superconducting fluctuations $f$ decreases linearly with volume until the $f = 0$ limit is reached at 426 $Å^3$ (where $MR = MR_a$), and above which $MR$ is almost volume-independent. The strong dependence of $f$ on cell volume is probably not a direct effect of decreasing carrier density, as volume only changes by ~1% across the linear regime, but more likely reflects structural tuning of the $CuO_2$ planes parameterised by cell volume. Structure refinements for $R =$ Gd and Sm materials [13], and for $R =$ Nd in ref. 6 show that Cu-O-Cu bond angle increases towards 180° with decreasing cell volume, optimising pair-formation through flattening of the $CuO_2$ planes.

Further systematic variations that evidence intrinsic inhomogeneous behaviour are seen when $MR_{7T, 4K}$ is plotted against doping $p$ for the $RuSr_2R_{1.8-x}Y_{0.2}Ce_xCu_2O_{10-\delta}$ series ($R =$ Nd, Sm and Gd). The striking result shown in Fig. 4b is that the $MR_{7T,4K}$ values for the three series converge towards a common point just below $p = 0.01$. This is not because their cell volumes converge – the volume for each series is approximately constant [13]. Instead convergence implies that the maximum possible concentration of fluctuations falls to $c_{max}(p) = 0$ at a critical doping $p = p_c$ which marks the transition at which superconducting fluctuations emerge in all the 1222 series. The $p_c = 0.0084 \pm 0.0002$ value estimated from the crossover of the three extrapolated $MR(p)$ variations is almost field-independent, as shown in the inset to Fig. 4b.

The idea that superconducting fluctuations emerge at $p_c$ is supported by fits to the $MR(p)$ variations using Eq. (1). The maximum concentration of fluctuations is written as $c_{max}(p) = (p-p_c)v$, where $(p-p_c)$ is the number of holes available for pairing in excess of the critical value, and $v = (V_{sf}/V_a)$ is the ratio of volumes occupied by a hole-carrier in a superconducting pair fluctuation, $V_{sf}$, and in the antiferromagnetic phase, $V_a$. The doping variation of the magnetoresistance for the antiferromagnetic matrix in the absence of pair fluctuations $MR_a(p)$ is needed to predict the $MR(p)$ curves. We have assumed a simple function $MR_a(p) = \exp(-p/P) - 1$ which has the correct $MR_a(p \gg p_c) = -1$ limit and fits the two known $MR_a$ values (at the $p = p_c$ convergence in Fig. 4b, and at the $c = 0$ limit of the $p = 0.03$ series in Fig. 4a) for constant $P = 0.057$. Using eq. (1) with this $MR_a(p)$ function accounts for the low-$p$ $MR_{7T,4K}$ variations for the $R$ = Nd, Sm and Gd series, assuming that each has an approximately constant value of $fv$ as shown on Fig. 4b ($f$ is dependent on cell volume as described above). The $MR$ data deviate above these curves at higher dopings as coherence between superconducting fluctuations is established, and $MR$ diverges at the onset of bulk superconductivity.

An approximate coherence length for the superconducting fluctuations is estimated by assuming maximum pairing ($f \approx 1$) for the Gd series of materials which have the highest observed value of $fv$, giving volume ratio $v = V_{sf}/V_a \approx 12.7$. The volume for each doped hole in the antiferromagnetic phase is $V_a \approx 105$ Å$^3$, as there are four Cu sites in each unit cell, so the effective volume of a superconducting pair fluctuation is estimated as $2V_{sf} \approx 2700$ Å$^3$. This equals the pair coherence volume in the limit that individual fluctuations are well separated, giving the mean coherence length for the pair as $\xi \approx 14$ Å at 4 K. This is in good agreement with typical $\xi = 5\text{-}15$ Å values obtained from conventional magnetic analyses of bulk cuprate superconductors.

The above results demonstrate that small variable concentrations of superconducting fluctuations within an insulating antiferromagnetic matrix are responsible for the large *MR* variations across low-doped 1222 ruthenocuprates. The systematic and simple variations of the low-temperature pair fluctuation density with cell volume and $p$ demonstrate that this aspect of the electrical inhomogeneity is intrinsic. It most likely reflects the atomistic method of doping cuprates; cell-to-cell differences in *R*/Ce-site cation or oxide vacancy content in the 1222 materials are influential because of the small coherence length for cuprate superconductivity. Phase separation or other microstructural effects are not a plausible explanation as the volume correlation in Fig. 4a extends over many 1222 materials from several different compositional systems.

The most significant discovery is the emergence transition for superconducting fluctuations at a small critical doping level $p_c$ = 0.0084, deep within the antiferromagnetic region of the cuprate phase diagram. This transition is distinct from the suppression of spontaneous copper spin antiferromagnetism, previously located at $p$ = 0.022 in the RuSr$_2$Nd$_{1.8-x}$Y$_{0.2}$Ce$_x$Cu$_2$O$_{10-\delta}$ system [6]. A similar emergent transition is expected in other cuprates but it may be difficult to detect. Although the antiferromagnetic regime of La$_{2-x}$Sr$_x$CuO$_4$ was not explored in a recent magnetoresistance study [1], we note that the fluctuation depairing fields reported from torque magnetometry measurements extrapolate linearly to zero at $p \approx$ 0.01 [18] providing plausible corroboration of the emergent transition.

In conclusion, this study shows that superconducting fluctuations emerge at a well-defined critical doping transition within the antiferromagnetic phase of cuprates. The low temperature density of superconducting fluctuations varies systematically across the antiferromagnetic and inhomogeneous regions, confirming the intrinsic nature of the electronic inhomogeneity and providing strong support for bosonic models where superconducting regions down to the size of individual pairs emerge within the antiferromagnetic phase.

This work was supported by EPSRC, the Royal Society and the Leverhulme Trust. We thank Prof. Nigel Hussey, University of Bristol, for comments on a draft manuscript.

Emails of corresponding authors: a.c.mclaughlin@abdn.ac.uk and j.p.attfield@ed.ac.uk

**Fig. 1** Temperature evolution of magnetoresistance for $RuSr_2R_{1.8-x}Y_{0.2}Ce_xCu_2O_{10-\delta}$ materials in a 5 T applied field; (a) R = Gd materials, showing hole doping levels $p$, and (b) R = Nd, Sm, Eu and Gd samples with $x = 0.9$ and constant doping $p = 0.03$. Discontinuities signifying Cu and Ru spin ordering transitions $T_{Cu}$ and $T_{Ru}$ are marked. The inset to (a) shows arrowed MR upturns at the onset of bulk superconductivity for $p \geq 0.041$, and at the upper temperatures for superconducting fluctuations $T_{sf}$ in samples with $p \leq 0.03$. The temperature derivative of $MR_{5T}$ for the R = Nd sample in the inset to (b) evidences a small density of superconducting fluctuations below ~30 K.

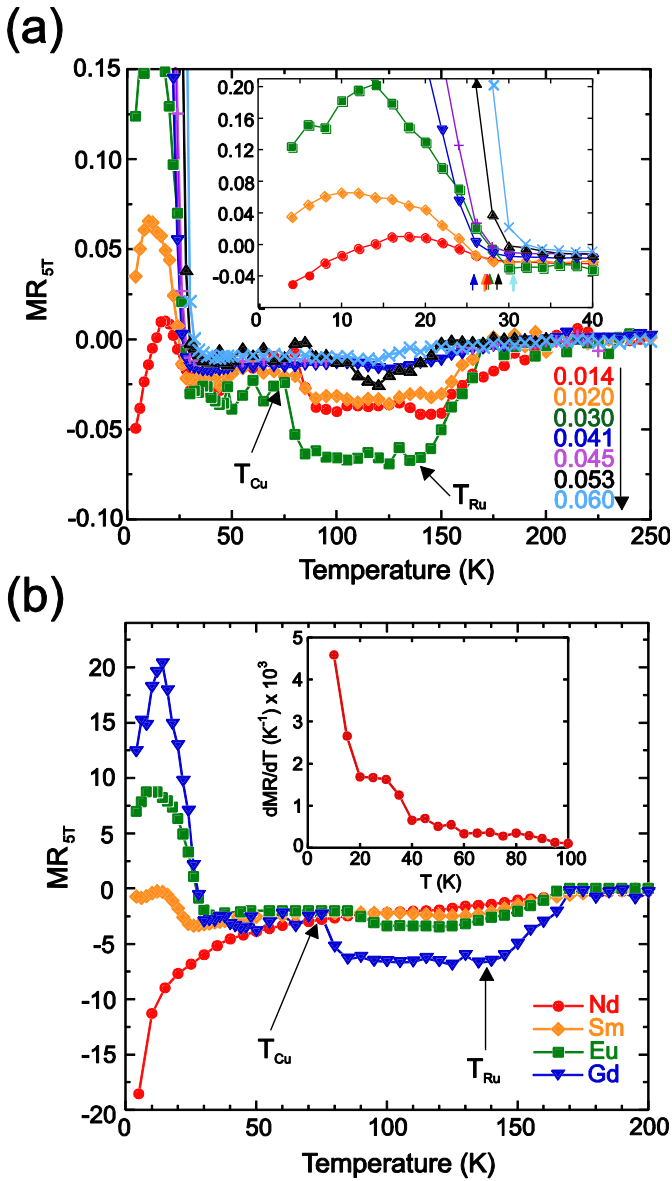

**Fig. 2** Electronic phase diagrams for RuSr$_2$R$_{1.8-x}$Y$_{0.2}$Ce$_x$Cu$_2$O$_{10-\delta}$ (R = Sm and Gd) derived from transitions in their $MR_{5T}(T)$ data, showing stability regions for the canted-antiferromagnetic order of Ru spins (c-AF(Ru)) below $T_{Ru}$ (filled circles), with additional antiferromagnetic Cu spin order (AF(Cu)) below $T_{Cu}$ (filled squares) or superconductivity (SC) below zero resistance $T_c$ (closed diamonds) at low temperatures. Superconducting fluctuations (SF) are observed below $T_{sf}$ (open diamonds) which starts at the $p_c = 0.0084$ critical point.

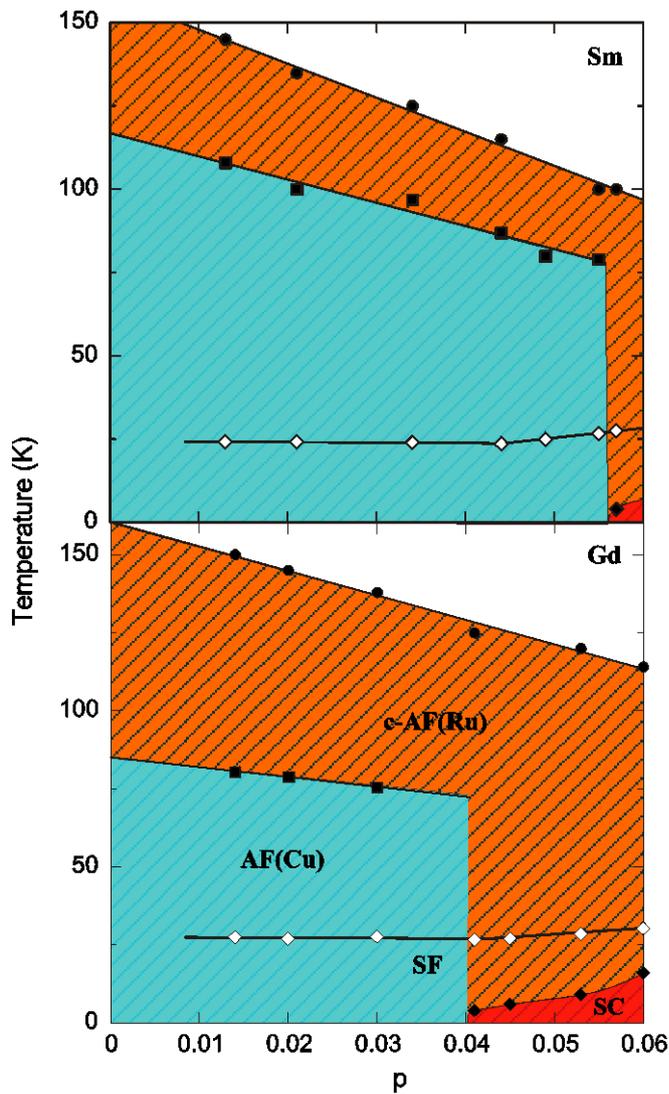

**Fig. 3.** Magnetoresistance variations for $RuSr_2R_{1.8-x}Y_{0.2}Ce_xCu_2O_{10-\delta}$ materials at 4 K. (a) Field variations for R = Sm samples labelled by doping levels $p$. The $p$ = 0.010 and 0.013 variations are fitted well as $MR(H) = -aH + bH^2$ with positive coefficients $a$ and $b$, characteristic of a doped antiferromagnetic insulator. (b) Variations of $MR/H$ at 4 K with doping for the R = Gd, Sm and Nd series. Low-doped samples have $(MR/H)_{7T} > (MR/H)_{2.5T}$ but this changes to $(MR/H)_{7T} < (MR/H)_{2.5T}$ as the influence of superconducting fluctuations increases with $p$ with the crossover at $p$ = 0.02 in the Sm series. The evolution of $(MR/H)$ confirms that superconducting fluctuations grow in the Nd series for $p > 0.03$ although no bulk superconductor is observed up to the highest available $p$ = 0.06 doping level [6].

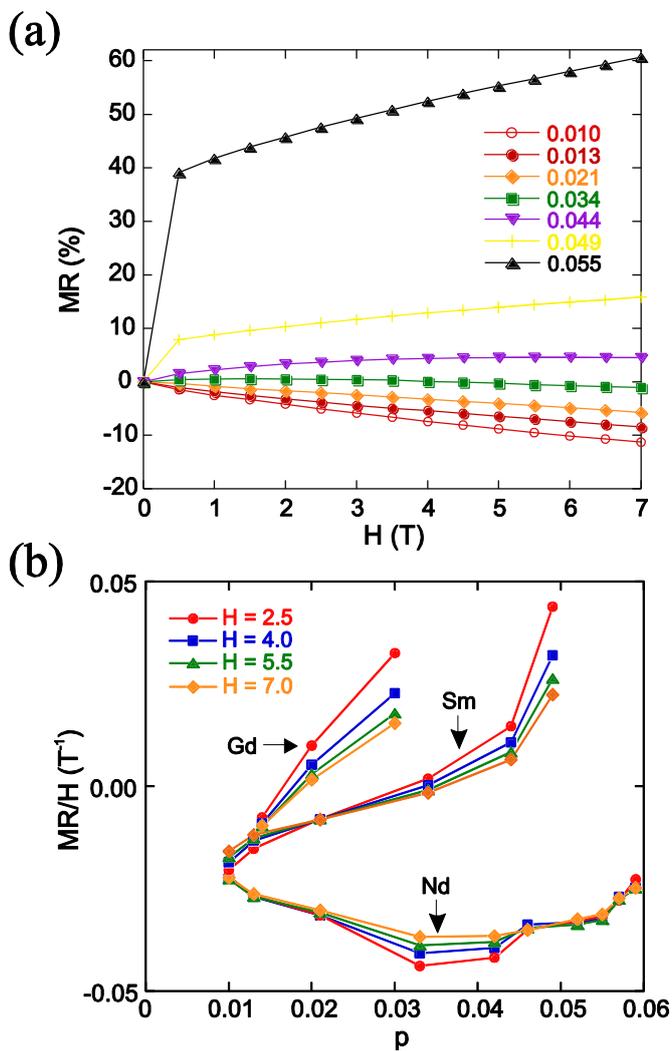

**Fig. 4** Magnetoresistance variations at a constant high field (7 T) and low temperature (4 K) for 1222 ruthenocuprates. (a) Plot of $MR_{7T,4K}$ against unit cell volume for 15 samples with fixed $p = 0.03$ (materials are from different chemical systems as shown in Supplementary Information). (b) Variations of $MR_{7T,4K}$ with doping $p$ for $RuSr_2R_{1.8-x}Y_{0.2}Ce_xCu_2O_{10-\delta}$ series with R = Nd, Sm, and Gd, also showing other $p = 0.03$ samples with linear volume-dependence in (a) within the box. The fits to $MR(p)$ are predicted from equation (1) for $fc_{max}(p) = (p-p_c)fv$ with $fv$ values for the three R series as shown, and converge at critical doping $p_c$. The $fv = 0$ line corresponds to the magnetoresistance of the antiferromagnetic matrix in the absence of pair fluctuations, $MR_a(p) = \exp(-p/0.057) - 1$. $MR$ data deviate above the curves at higher dopings ($p > 0.03$) where coherence between superconducting fluctuations is established, and $MR$ diverges at the onset of bulk superconductivity shown by arrows for the Gd and Sm series. The inset shows the field-variation of the critical doping at 4 K; $p_c$ could not be estimated accurately for fields below 3 T.

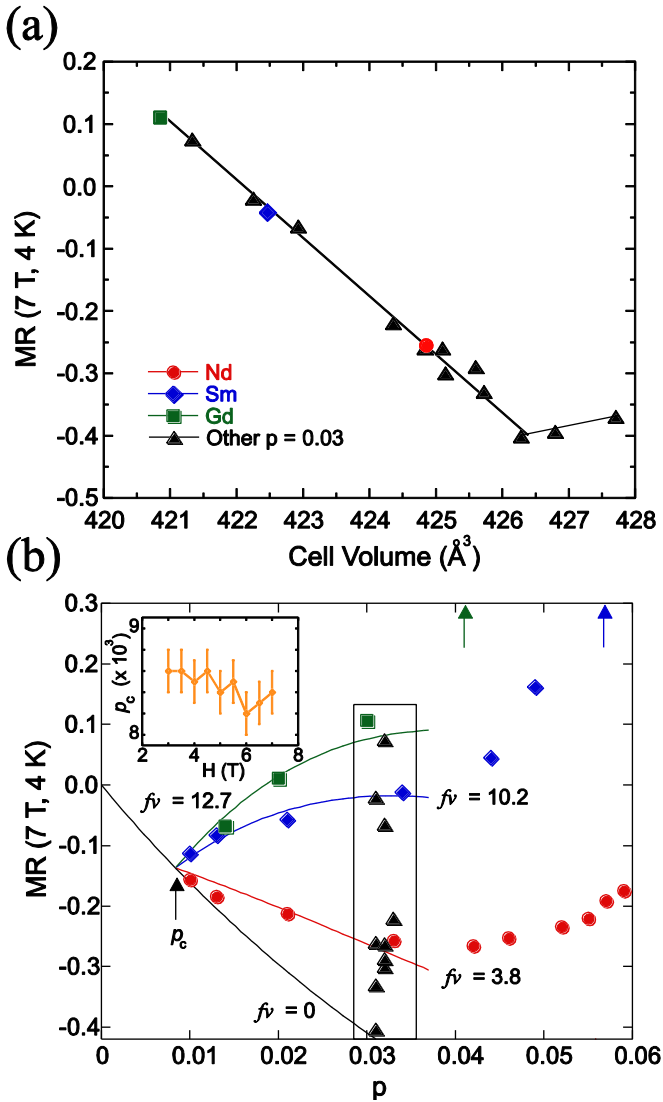